\documentclass[12pt]{article}

\usepackage{amsmath}
\usepackage{epsf}

\allowdisplaybreaks

\begin{document}

\baselineskip=18.2pt plus 0.2pt minus 0.1pt

\makeatletter
\@addtoreset{equation}{section}
\renewcommand{\theequation}{\thesection.\arabic{equation}}
\renewcommand{\thefootnote}{\fnsymbol{footnote}}
\newcommand{\nn}{\nonumber}
\newcommand{\wt}[1]{\widetilde{#1}}
\newcommand{\bra}[1]{\langle #1\vert}
\newcommand{\ket}[1]{\vert #1\rangle}
\newcommand{\braket}[2]{\langle #1\vert #2\rangle}
\newcommand{\bbbk}[4]{{}_1\langle #1|{}_2\langle #2|
                      {}_3\langle #3|#4\rangle_{123}}
\newcommand{\delB}{\delta_{\text B}}
\newcommand{\QB}{Q_{\text B}}
\newcommand{\N}[2]{N^{#1}_{#2}}
\newcommand{\X}[2]{X^{#1}_{#2}}
\newcommand{\V}[2]{V^{#1}_{#2}}
\newcommand{\W}[2]{\widetilde{V}^{#1}_{#2}}
\newcommand{\p}{\partial}
\newcommand{\ap}{\alpha'}
\newcommand{\vac}{\ket{0}}
\newcommand{\intund}[1]{\int\nolimits_{#1}}
\newcommand{\Ngh}{N_{\rm gh}}
\newcommand{\Nzz}{N_{00}}
\newcommand{\bm}[1]{\boldsymbol{#1}}
\newcommand{\calNc}{{\cal N}_{\rm c}}
\newcommand{\calV}{{\cal V}}
\newcommand{\calC}{{\cal C}}
\newcommand{\calQ}{{\cal Q}}
\newcommand{\calO}{{\cal O}}
\newcommand{\Pmatrix}[1]{\begin{pmatrix} #1 \end{pmatrix}}
\newcommand{\SV}{{\cal S}}
\newcommand{\f}{f}
\newcommand{\phic}{\phi_{\rm c}}
\newcommand{\Psic}{\Psi_{\rm c}}
\newcommand{\calQB}{{\cal Q}_{\rm B}}
\newcommand{\Phit}{\Phi_{\rm t}}
\newcommand{\phit}{\phi_{\rm t}}
\newcommand{\calNt}{{\cal N}_{\rm t}}
\newcommand{\calS}{{\cal S}_{\rm T}}
\newcommand{\calM}{{\cal M}}
\newcommand{\vpz}{\bm{v}_+ -\bm{v}_0}
\newcommand{\vmz}{\bm{v}_- -\bm{v}_0}
\newcommand{\vpm}{\bm{v}_+ -\bm{v}_-}
\newcommand{\vmp}{\bm{v}_- -\bm{v}_+}
\newcommand{\go}{g_{\rm o}}
\newcommand{\mt}{m_{\rm t}}
\newcommand{\varphit}{\varphi_{\rm t}}
\newcommand{\vvpz}{\bm{v}_{+0}}
\newcommand{\vvmz}{\bm{v}_{-0}}
\newcommand{\T}{{\rm T}}
\newcommand{\calEc}{{\cal E}_{\rm c}}
\newcommand{\Zmatt}{Z_{\rm matt}}
\newcommand{\Zgh}{Z_{\rm gh}}
\newcommand{\gh}{{\rm g}}
\newcommand{\matt}{{\rm matt}}
\newcommand{\diag}{\mathop{\rm diag}}
\newcommand{\Psip}{\Phi_{\rm v}}
\newcommand{\Phip}{\Phi_{\rm v}}
\newcommand{\phip}{\phi_{\rm v}}

\begin{titlepage}
\title{
\hfill\parbox{4cm}
{\normalsize KUNS-1732\\UT-963\\{\tt hep-th/0108150}}\\
\vspace{1cm}
{\Large\bf
Open String States around a Classical Solution in
Vacuum String Field Theory
}
}
\author{
Hiroyuki {\sc Hata}\thanks{{\tt hata@gauge.scphys.kyoto-u.ac.jp}}
\\[7pt]
{\it Department of Physics, Kyoto University, Kyoto 606-8502, Japan}
\\[10pt]
and
\\[10pt]
Teruhiko {\sc Kawano}\thanks{
{\tt kawano@hep-th.phys.s.u-tokyo.ac.jp}}
\\[7pt]
{\it Department of Physics, University of Tokyo}\\
{\it Hongo, Tokyo 113-0033, Japan}
}
\date{\normalsize August, 2001}
\maketitle
\thispagestyle{empty}
\begin{abstract}
\normalsize\noindent
We construct a classical solution of vacuum string field theory (VSFT)
and study whether it represents the perturbative open string vacuum.
Our solution is given as a squeezed state in the Siegel gauge, and it
fixes the arbitrary coefficients in the BRST operator in VSFT.
We identify the tachyon and massless vector states as
fluctuation modes around the classical solution.
The tachyon mass squared $\ap\mt^2$ is given in a closed form using
the Neumann coefficients defining the three-string vertex, and
it reproduces numerically the expected value of $-1$ to high
precision. The ratio of the potential height of the solution to
the D25-brane tension is also given in terms of the Neumann
coefficients. However, the behavior of the potential height in level
truncation does not match our expectation, though there are subtle
points in the analysis.

\end{abstract}

\end{titlepage}

\section{Introduction}

String field theory has proved to be a powerful tool in
understanding the conjectures about tachyon condensation in bosonic
open string theory \cite{Sen:1999mh,Sen:1999mg,Sen:1999xm}.
Among the three conjectures, the potential height problem and the
descent relation have been fully understood
(see \cite{Ohmori:2001am} and the references therein), and the current
interest has now been focused on the third problem: Is the pure closed
string theory without physical open string excitations realized at
the tachyon vacuum?

There have appeared a number of works
\cite{HataTera,Ellwood:2001py,Ellwood:2001ig}
which studied this problem in
the level truncation approximation in cubic open string field theory
(CSFT) and obtained results supporting the absence of open string
excitations at the tachyon vacuum. However, in contrast to the case of
the potential height problem \cite{SZ,MT}, the level truncation cannot
give a conclusive answer to the third problem since we have to deal
with the space of open string states with level number extending to
infinity.
For a complete understanding of the third problem, we would
need the exact solution of tachyon vacuum in CSFT.

Vacuum string field theory (VSFT)
\cite{VSFT1,VSFT2,VSFT3,VSFTR} has been proposed to study the third
problem in a reverse way. It is an open string field theory
which contains no physical open string excitations at all and hence is
expected to describe the tachyon vacuum of ordinary CSFT.
The action of VSFT is the same as that of CSFT except that the BRST
operator $\QB$ is replaced with another one $\calQ$ consisting solely
of the ghost coordinate. Due to this pure ghost structure of the new
BRST operator $\calQ$, physical open string spectrum of VSFT becomes
trivial.

However, for VSFT to solve the third problem,
we have to show that it is connected with the perturbative open
string theory.\footnote{
The descent relation has been studied in VSFT to give expected
result under the assumption that the ghost parts of the classical
solutions for D$p$-branes with different $p$ are common
\cite{VSFT1,VSFT2}.
}
Namely, we have to show that VSFT has a classical
solution (let us denote it $\Psic$) describing the perturbative open
string vacuum. Concretely,  $\Psic$ has to satisfy the following two:
First, the physical fluctuation spectrum around $\Psic$ must
reproduce that of the perturbative open string theory.
Second, the raise of the energy density of the $\Psic$ state from
that of the trivial vacuum in VSFT must be equal to the D25-brane
tension.
Elaborated reformulations of VSFT developed recently
\cite{Half,GT1,KawaOku,GT2,FuruOku} would be useful for studying these
problems.

The purpose of the present paper is to construct an exact classical
solution of VSFT and study whether this solution satisfies the above
two requirements.
By taking the Siegel gauge, the classical solution has to satisfy two
equations, one is the part of $\calQ\Psic + \Psic *\Psic=0$
not proportional to the anti-ghost zero-mode $b_0$ and the other is
the part proportional to $b_0$ (the latter equation was called the
BRST invariance condition in \cite{HS}).
The former equation is easily solved by assuming the squeezed state
form for both the matter and the ghost part of the solution.
In fact, such solution including the ghost part has essentially been
constructed in \cite{KP}.
Moreover, it has been claimed \cite{VSFT2} that (the matter part of)
the solution is identical with the sliver state \cite{TSU} constructed
in a different manner as the matter part of a classical solution in
VSFT.\footnote{
An earlier attempt to construct VSFT solutions by considering ghost
number one excitations on the sliver state has appeared in
\cite{David}.
}
On the other hand, $\Psic$ must also solve the BRST invariance
condition. We find that this condition does not impose
further constraint on $\Psic$ but rather it fixes the coefficients
in $\calQ$ which were arbitrary at the start.
Namely, the arbitrary coefficients in $\calQ$ are uniquely determined
by the requirement that VSFT has a translationally and Lorentz
invariant solution in the Siegel gauge.

To test whether our classical solution $\Psic$ corresponds to the
perturbative open string vacuum, we first study the fluctuation
spectrum around $\Psic$. Namely, we consider the wave equation
$\calQB\Phi=0$ where $\calQB$ is the BRST operator for fluctuations
around $\Psic$.
As solutions to the wave equation, we construct a scalar and a vector
solution which we expect to represent the tachyon and the massless
vector states in perturbative open string theory.
For the scalar solution, we obtain from the wave equation
$\calQB\Phi=0$ an expression of mass squared $\ap\mt^2$ of the
scalar particle. It is given in a closed form using the Neumann
coefficients defining the three-string vertex.
Since we cannot evaluate this $\ap\mt^2$ analytically at present,
we calculate it numerically by using level truncation.
The result is just as we expected for the tachyon state:
$\ap\mt^2$ approaches to $-1$ to high precision as the level
cutoff is increased.

For the vector solution, our analysis is not complete.
Though we can show that the solution represents an exactly massless
state, the transversality condition for the polarization vector
is not imposed by the wave equation. Therefore, more detailed
analysis of the solution space including the ghost mode is necessary
for the massless sector.
As a test of our vector solution we also calculate the
tachyon-tachyon-vector coupling to find that it agrees with a
familiar one having gauge invariance.

After establishing the tachyon wave function, we proceed to the
test of the potential height.
For this purpose, we have to identify the D25-brane tension $T_{25}$
and hence the open string coupling constant $\go$. For calculating
$\go$, we first determine the normalization of the tachyon wave
function so that the VSFT action reproduces the canonical kinetic term
for the tachyon field. Then, $\go$ is given as the three tachyon
coupling on the mass-shell.
After these preparations, we obtain the ratio of the energy density
$\calEc$ of the $\Psic$ state to the tension $T_{25}$. It is given in
a closed form in terms of the Neumann coefficients. In particular, the
determinant factor appearing in the energy density $\calEc$ is
cancelled with that in $T_{25}$.
We calculate the ratio $\calEc/T_{25}$ numerically using level
truncation. However, the result is an unwelcome one:
$\calEc/T_{25}$ becomes far beyond the expected value of one as the
level number cutoff is increased. Though this result is apparently
disappointing, we find that there are ambiguities in the analysis of
$\calEc/T_{25}$ which need to be fixed before getting a conclusive
answer to the potential height problem.

The organization of the rest of this paper is as follows.
In sec.\ 2, we summarize the VSFT action and various properties of the
Neumann coefficients used in later sections.
In sec.\ 3, the classical solution $\Psic$ of VSFT is constructed
and the coefficients in the BRST operator of VSFT is fixed.
In sec.\ 4, we construct the tachyon and massless vector wave
functions as fluctuation modes around $\Psic$, and analyze the tachyon
mass. In sec.\ 5, we study the potential height of our classical
solution. In the final section, we conclude the paper and discuss
open questions. In the appendix, we present some useful formulas
and technical details of the calculations.

\section{VSFT action}

We shall consider the VSFT described by the following action
\cite{VSFT1,VSFT2,VSFTR}:
\begin{align}
\SV[\Psi]&=-K\left(
\frac12 \Psi\cdot\calQ\,\Psi+\frac13\Psi\cdot(\Psi * \Psi)\right)
\nn\\
&= -K\left(
\frac12\intund{b_0,x}\bra{\Psi}\calQ\ket{\Psi}+ \frac13
\intund{b_0,x}^{(3)}\intund{b_0,x}^{(2)}\intund{b_0,x}^{(1)}
\bbbk{\Psi}{\Psi}{\Psi}{V}\right) .
\label{eq:SV}
\end{align}
Here, we are taking the representation of diagonalizing the anti-ghost
zero-mode $b_0$, and
$\intund{b_0,x}^{(r)}\equiv\int db_0^{(r)}\int d^{26}x_r$
denotes the integration over $b_0$ and the
center-of-mass coordinate $x^\mu$ of the $r$-th string.
The string field $\ket{\Psi(x,b_0)}$ is a state
in the Fock space of the first quantized string and carries ghost
number $-1$.
The three-string vertex $\ket{V}$ is the same as in the ordinary CSFT
and is given in the momentum representation for the center-of-mass
$x^\mu$ as \cite{CST,Samuel,Ohta,GJI,GJII,IOS}
\begin{align}
\ket{V}_{123}
&=\exp\left\{
\sum_{r,s=1}^3 \left(
-\sum_{n,m\ge 0}\frac12 a^{(r)\dagger}_n \V{rs}{nm} a^{(s)\dagger}_m
+ \sum_{n\ge 1,m\ge 0}c^{(r)\dagger}_n \W{rs}{nm} b^{(s)\dagger}_m
\right)\right\}\!\ket{0}_{123}
\nn\\
&\hspace*{5cm}\times (2\pi)^{26}\delta^{26}\left(p_1+p_2+p_3\right) ,
\label{eq:V}
\end{align}
where $\V{rs}{nm}$ and $\W{rs}{nm}$ are the Neumann coefficients for
the symmetric three-string connection.\footnote{
These Neumann coefficients are related to the ones, $\N{rs}{nm}$ and
$\X{rs}{nm}$, in refs.\ \cite{T,HS} by
$\V{rs}{nm}=-\sqrt{n}\N{rs}{nm}\sqrt{m}$ ($n,m\ge 1$),
$\V{rs}{n0}=-\sqrt{n}\N{rs}{n0}$ ($n\ge 1$),
$\V{rs}{00}=-\N{rs}{00}$,
and $\W{rs}{nm}=-\X{rs}{nm}$ ($n\ge 1, m\ge 0$).
}
The matter (ghost) oscillators satisfy the following
(anti-)commutation relations
\begin{equation}
[a^{(r)\mu}_n, a^{(s)\nu\dagger}_m]=\eta^{\mu\nu}\delta_{nm}\delta^{rs},
\quad
\{c^{(r)}_n,b^{(s)\dagger}_m\} = \{b^{(r)}_n,c^{(s)\dagger}_m\}
=\delta_{nm}\delta^{rs} ,
\quad (n,m\ge 1) ,
\end{equation}
and the oscillator vacuum $\ket{0}$ is defined by $(a_n,b_n,c_n)\vac=0$
($n\ge 1$).
As for the zero-modes, we have $b^{\dagger (r)}_0= b^{(r)}_0$,
$c^{\dagger (r)}_0= c^{(r)}_0=\p/\p b_0^{(r)}$
and $a^{(r)}_0 =a^{\dagger(r)}_0= \sqrt{2}\,p_r$
($p_r=-i\p/\p x_r$ is the center-of-mass momentum of the string $r$,
and we are adopting the convention of $\ap=1$).

The essential difference of the VSFT action from that of the
ordinary CSFT is the BRST operator $\calQ$ around the tachyon vacuum.
It consists purely of ghost operators:
\begin{equation}
\calQ = c_0 + \sum_{n\ge 1}\f_n \calC_n ,
\quad
\calC_n \equiv  c_n + (-1)^n c_n^\dagger ,
\label{eq:calQ}
\end{equation}
where the coefficient $f_n$ is real (pure imaginary) for an even (odd)
$n$ due to the requirement that $\calQ$ be hermitian, but otherwise
arbitrary at the present stage.
Namely, $\calQ$ satisfies the nilpotency and the Leibniz rule on the
$*$-product for arbitrary $f_n$,
and the VSFT action (\ref{eq:SV}) has  an invariance under the gauge
transformation
\begin{equation}
\delta_\Lambda\Psi = \calQ\Lambda
+ \Psi *\Lambda - \Lambda *\Psi .
\label{eq:gaugetransf}
\end{equation}
Since the cohomology of $\calQ$ is trivial, the quadratic term
$-K(1/2)\Psi\!\cdot\!\calQ\Psi$ of the VSFT action (\ref{eq:SV})
supports no physical open string excitations at all.

Here, we shall mention the hermiticity constraint on the string field.
The string field $\Psi$ is restricted to satisfy the following
hermiticity condition:
\begin{equation}
{}_2\bra{\Psi}
=\intund{b_0,x}^{(1)}
{}_{12}\braket{R}{\Psi}_1 ,
\label{eq:hermiticity}
\end{equation}
where ${}_{12}\bra{R}$ is the reflector given in the momentum
representation as
\begin{align}
{}_{12}\bra{R} &= {}_{12}\bra{0}\exp\left\{
-\sum_{n\ge 1}(-1)^n\Bigl(a^{(1)}_n a^{(2)}_n
+ c^{(1)}_n b^{(2)}_n + c^{(2)}_n b^{(1)}_n \Bigr)\right\}
\nn\\
&\qquad\qquad
\times(2\pi)^{26}\delta^{26}(p_1+p_2)\,\delta(b_0^{(1)}-b_0^{(2)}) .
\end{align}
This constraint reduces the number of degrees of freedom in $\Psi$ to
half and ensures the hermiticity of the action (\ref{eq:SV}).

In the rest of this section, we shall summarize various useful
properties of the Neumann coefficients appearing in the vertex
(\ref{eq:V}).
Due to the cyclic symmetry property for the three strings, the Neumann
coefficients have only three independent components with respect the
the upper indices, and we define
\begin{equation}
(V_0)_{nm}=\V{rr}{nm},\qquad
(V_\pm)_{nm}=\V{r,r\pm 1}{nm},\quad
\label{eq:V0pm}
\end{equation}
for $n, m\ge 1$.
Then, the twist transformation property of the vertex,
\begin{equation}
\Omega_1\Omega_2\Omega_3\ket{V}_{123}
=\ket{V}_{321} ,
\label{eq:OmegaOmegaOmegaV=V}
\end{equation}
is translated to the following for the Neumann coefficients:
\begin{equation}
C V_0 C =V_0,\qquad C V_\pm C =V_\mp .
\label{eq:CVC=V}
\end{equation}
Here, $\Omega_r$ is the twist operator on the Fock space of the string
$r$:
\begin{equation}
\Omega\left(a_n,b_n,c_n\right)\Omega^{-1}
=(-1)^n\left(a_n,b_n,c_n\right) ,
\quad
\Omega\ket{0}=\ket{0} ,
\label{eq:Omega}
\end{equation}
and $C$ is the twist matrix $C$ defined by
\begin{equation}
C_{nm} =(-1)^n\delta_{nm} ,
\qquad (n,m\ge 1) .
\label{eq:C}
\end{equation}

Next, let us define the matrices $M_0$, $M_+$ and $M_-$ by
\begin{equation}
M_0=C V_0,\qquad M_\pm =C V_\pm .
\label{eq:M}
\end{equation}
They enjoy the following two basic properties:

\noindent
(i) $M_\alpha$ ($\alpha=0,\pm$) are commutative to each other:
\begin{equation}
[M_0,M_\pm]=[M_+,M_-]=0 .
\label{eq:[M,M]=0}
\end{equation}

\noindent
(ii) $M_\alpha$ satisfy the two identities:
\begin{align}
&M_0 + M_+ + M_- = 1 ,
\label{eq:M+M+M=1}
\\
&M_+ M_- = M_0^2 - M_0 .
\label{eq:MM=M^2-M}
\end{align}

\noindent
The following formulas are consequences of (\ref{eq:M+M+M=1}) and
(\ref{eq:MM=M^2-M}):
\begin{align}
&M_0^2 +M_+^2 +M_-^2 =1 ,
\label{eq:M^2+M^2+M^2=1}
\\
&M_+^3 +M_-^3 = 1 - 3M_0^2 + 2M_0^3 ,
\label{eq:M^3+M^3=}
\\
&M_\pm^2 - M_\pm = M_0 M_\mp .
\label{eq:M^2-M=MM}
\end{align}
We also have the corresponding equations for the ghost Neumann
coefficients: eqs.\ (\ref{eq:V0pm}) -- (\ref{eq:M^2-M=MM}) with $V$ and
$M$ replaced by $\wt{V}$ and $\wt{M}$, respectively.

Next are the formulas concerning the Neumann coefficients
$V_{n0}^{rs}$.
Let us define the vectors $\bm{v}_0$ and $\bm{v}_\pm$ by
\begin{equation}
\left(\bm{v}_0\right)_n= \V{rr}{n0},
\qquad
\left(\bm{v}_\pm\right)_n= \V{r,r\pm 1}{n0},
\qquad (n\ge 1) .
\label{eq:vecv0pm}
\end{equation}
Under the twist transformation, we have
\begin{equation}
C\bm{v}_0 = \bm{v}_0,\qquad
C\bm{v}_\pm = \bm{v}_\mp .
\label{eq:Cv=v}
\end{equation}
Now, the following equations have been known to hold \cite{GJI,GJII}:
\begin{align}
&\sum_{t=1}^3\sum_{n\ge 1}\V{rt}{mn}\V{ts}{n0}=\V{rs}{m0} \ ,
\qquad (m \ge 1)
\label{eq:VV0=V0}\\
&\sum_{t=1}^3\sum_{n\ge 1}\V{tr}{n0}\V{ts}{n0}
= 2\,\V{rs}{00} .
\label{eq:V0V0=V00}
\end{align}
Here, we must bear in mind that these equations are valid only when
the upper open indices associated with the zero-mode ($s$ in
(\ref{eq:VV0=V0}) and $r$ and $s$ in (\ref{eq:V0V0=V00}))
are contracted with conserved quantities.
Using this fact and (\ref{eq:M+M+M=1}), eq.\ (\ref{eq:VV0=V0}) is
reexpressed as
\begin{align}
&M_+ \vvmz + M_- \vvpz=0 ,
\label{eq:MvI}
\\
&M_0 \bm{v}_{\pm 0}+(M_\mp-1)\bm{v}_{\mp 0}=0 ,
\label{eq:MvII}
\end{align}
where we have introduced abbreviated notations:
\begin{equation}
\bm{v}_{\pm 0}=\bm{v}_\pm - \bm{v}_0 .
\label{eq:vvpmz}
\end{equation}
On the other hand, (\ref{eq:V0V0=V00}) together with
\begin{equation}
\V{rs}{00}= V_{00}\,\delta^{rs} ,
\qquad
V_{00}= \frac12 \ln\left(\frac{3^3}{2^4}\right) ,
\label{eq:V00}
\end{equation}
and (\ref{eq:Cv=v}) gives
\begin{equation}
2\,(\vvpz)^2 - \vvpz\!\cdot\!\vvmz=2\,V_{00} .
\label{eq:vv=V00}
\end{equation}
We do not have the corresponding equations to
(\ref{eq:MvI}), (\ref{eq:MvII}) and (\ref{eq:vv=V00}) for the ghost
Neumann coefficients.

\section{Classical solution in the Siegel gauge}

We would like to construct a (translationally and Lorentz invariant)
classical solution $\Psic$ to the equation of motion of the VSFT
action (\ref{eq:SV}),
$\calQ \Psic + \Psic\! *\! \Psic =0$, which is expressed in the
Fock space representation as
\begin{equation}
\calQ\ket{\Psic}_3 +
\intund{b_0,x}^{(2)}\intund{b_0,x}^{(1)}
{}_1\bra{\Psic}{}_2\braket{\Psic}{V}_{123}=0 .
\label{eq:eqmotSV}
\end{equation}
It is the task of later sections to examine whether this
solution represents the perturbative open string vacuum.

Due to the pure ghost form of $\calQ$, the equation of motion
(\ref{eq:eqmotSV}) can be solved explicitly by adopting the Siegel
gauge for the solution, $b_0\Psic=0$.
Substituting the expression
\begin{equation}
\ket{\Psic}=b_0\ket{\phic} ,
\label{eq:Psic=b0phic}
\end{equation}
into (\ref{eq:eqmotSV}), we find that it consists of the following two
equations for $\ket{\phic}$:
\begin{align}
&\ket{\phic}_3 + {}_1\bra{\phic}{}_2
\braket{\phic}{\widehat{V}}_{123}\Bigr\vert_{p_r=0}
=0 ,
\label{eq:delpot=0}
\\
&\sum_{n \ge 1}\f_n\calC_n\ket{\phic}_3
+ {}_1\bra{\phic}{}_2\bra{\phic}\sum_{r=1}^3\sum_{n\ge 1}
c^{\dagger (r)}_n \W{r3}{n0}
\ket{\widehat{V}}_{123}\Bigr\vert_{p_r=0}=0 ,
\label{eq:BRSTinv}
\end{align}
where $\ket{\widehat{V}}$ is the reduced vertex without $b_0$:
\begin{align}
\ket{\widehat{V}}_{123}
&=\exp\left\{
\sum_{r,s=1}^3 \left(
-\sum_{n,m\ge 0}\frac12 a^{(r)\dagger}_n \V{rs}{nm} a^{(s)\dagger}_m
+ \sum_{n,m\ge 1}c^{(r)\dagger}_n \W{rs}{nm} b^{(s)\dagger}_m
\right)\right\}\!\ket{0}_{123} .
\end{align}
The original equation of motion (\ref{eq:eqmotSV}) is given by
$\mbox{eq.\,}(\ref{eq:delpot=0}) - b_0^{(3)}\!\times\!
\mbox{eq.\,}(\ref{eq:BRSTinv})$.
If we start with the gauge-fixed action, (\ref{eq:SV}) with
(\ref{eq:Psic=b0phic}) substituted,
(\ref{eq:delpot=0}) is its equation of motion, while
(\ref{eq:BRSTinv}) is the BRST invariance condition \cite{HS}.

Let us consider the first equation (\ref{eq:delpot=0}).
It has been known \cite{KP,VSFT2} that (\ref{eq:delpot=0}) can be
solved (including the ghost part) by assuming the squeezed state form
for $\ket{\phic}$:\footnote{
However, our explicit formulas for the ghost part differ from
those in \cite{KP}.
}
\begin{equation}
\ket{\phic}=\calNc\exp\left(-\frac12 \sum_{n,m\ge 1}
a^\dagger_n S_{nm}a^\dagger_m + \sum_{n,m\ge 1}
c^\dagger_n \wt{S}_{nm} b^\dagger_m\right)\ket{0} ,
\label{eq:phic}
\end{equation}
where $S_{nm}$ and $\wt{S}_{nm}$ are unknown real
coefficients\footnote{
The hermiticity constraint (\ref{eq:hermiticity}) for $\Psic$ implies
that $\calO^*_{nm}=(-1)^{n+m}\calO_{nm}$ for both $\calO=S$ and
$\wt{S}$.
Imposing further the twist invariance condition (\ref{eq:twistinv}),
$S_{nm}$ and $\wt{S}_{nm}$ are restricted to be real.
}
and $\calNc$ is
the (real) normalization factor.
We assume further that the state $\ket{\phic}$ is twist invariant,
$\Omega\ket{\phic}=\ket{\phic}$, and hence $S_{nm}$ and $\wt{S}_{nm}$
satisfy the matrix equations
\begin{equation}
C S C = S, \qquad C \wt{S}C = \wt{S} .
\label{eq:twistinv}
\end{equation}

Formulas (\ref{eq:formulaB}) and (\ref{eq:formulaF}) in appendix
\ref{sec:formulas} imply that, for the squeezed state
$\ket{\phic}$ (\ref{eq:phic}), the second term of (\ref{eq:delpot=0})
is again a squeezed state, and moreover it becomes proportional to
$\ket{\phic}$,
\begin{equation}
{}_1\bra{\phic}{}_2\braket{\phic}{\widehat{V}}_{123}
\Bigr\vert_{p_r=0}=
\calNc \left[\det(1-S\calV)\right]^{-13}
\det(1-\wt{S}\wt{\calV})\,\ket{\phic}_3 ,
\label{eq:phiphiv=phi}
\end{equation}
provided $S$ and $\wt{S}$ satisfy
\begin{align}
&S=V_0 + (V_+,V_-)(1-S\calV)^{-1} S\Pmatrix{V_-\\ V_+} ,
\label{eq:eqforS}
\end{align}
and the same one with all the matrices replaced with the tilded ones,
respectively.
In (\ref{eq:eqforS}), $\calV$ is
\begin{equation}
\calV = \Pmatrix{V_0 & V_+\\ V_- & V_0} ,
\label{eq:calV}
\end{equation}
and $S$ on the RHS should read $\diag(S,S)$.
If (\ref{eq:eqforS}) and the corresponding one for $\wt{S}$ hold,
then $\ket{\phic}$ (\ref{eq:phic}) becomes a solution to
(\ref{eq:delpot=0}) by taking the following normalization factor
$\calNc$:
\begin{equation}
\calNc = - \left[\det(1-S\calV)\right]^{13}
[\det(1-\wt{S}\wt{\calV})]^{-1}
\label{eq:calNc}
\end{equation}

Eq.\ (\ref{eq:eqforS}) for $S$ has been solved in \cite{KP,VSFT2}, and
we shall summarize the points in obtaining the solution.
Defining
\begin{equation}
T=CS=SC ,
\label{eq:T=CS}
\end{equation}
eq.\ (\ref{eq:eqforS}) multiplied by $C$ on the left reads
\begin{equation}
T=M_0 + (M_+,M_-)(1-T\calM)^{-1}\,T\Pmatrix{M_-\\ M_+} ,
\label{eq:eqforT}
\end{equation}
with
\begin{equation}
\calM = \Pmatrix{M_0 & M_+\\ M_- & M_0} .
\label{eq:calM}
\end{equation}
Let us assume that $T$ commutes with the matrices $M_\alpha$:
\begin{equation}
[T,M_\alpha]=0, \qquad (\alpha=0,\pm) .
\label{eq:[T,M]=0}
\end{equation}
Then, all the matrices in (\ref{eq:eqforT}) are commutative (recall
(\ref{eq:[M,M]=0})), which makes (\ref{eq:eqforT}) fairly easy to
deal with.
Using the formulas (\ref{eq:M+M+M=1})--(\ref{eq:M^3+M^3=})
for $M_\alpha$ and, in particular,
\begin{equation}
(1-S\calV)^{-1}=\left(1-T\calM\right)^{-1}
=\left(1-2 M_0 T + M_0 T^2\right)^{-1}
\Pmatrix{1-T M_0 & T M_+\\
           T M_- & 1- T M_0} ,
\label{eq:(1-TM)^-1}
\end{equation}
eq.\ (\ref{eq:eqforT}) is reduced to
\begin{equation}
(T - 1)\Bigl(M_0\,T^2 -(1 + M_0)T+ M_0\Bigr)=0 .
\label{eq:(T-1)()=0}
\end{equation}
We do not adopt the solution $T=1$ which corresponds to the identity
state, and take a solution to
\begin{equation}
M_0\,T^2 -(1 + M_0)T+ M_0=0 .
\label{eq:M0T^2-(1+M0)T+M0=0}
\end{equation}
Among (infinitely) many solutions to (\ref{eq:M0T^2-(1+M0)T+M0=0})
we take the following one,
\begin{equation}
T=\frac{1}{2 M_0}\left(
1 + M_0 - \sqrt{(1-M_0)(1 + 3 M_0)}\right) ,
\label{eq:T-}
\end{equation}
where the branch of the matrix square root is defined by the Taylor
expansion,
\begin{equation}
\sqrt{(1-M_0)(1 + 3 M_0)}=1 + \sum_{k=1}^\infty
\Pmatrix{1/2\\ k}\left(2 M_0 - 3 M_0^2\right)^k ,
\label{eq:TaylorExp}
\end{equation}
and hence $T$ has an expansion in positive powers of $M_0$:
$T=M_0 - M_0^2 + 2 M_0^3 + \cdots$.
The commutativity (\ref{eq:[T,M]=0}) is evidently satisfied.
It has been claimed by numerical comparison that the matrix $S=CT$
with $T$ given by (\ref{eq:T-}) is identical to the matrix defining
the matter part of the sliver state \cite{VSFT2}.
However, we do not use this fact explicitly in the rest of this
paper.

Determination of the ghost part matrix $\wt{S}$ in (\ref{eq:phic}) is
exactly the same as for $S$. We take as $\wt{T}=C\wt{S}=\wt{S}C$ the
same one (\ref{eq:T-}) with $M_0$ replaced with $\wt{M}_0$:
\begin{equation}
\wt{T}=\frac{1}{2 \wt{M}_0}\left(
1 + \wt{M}_0 - \sqrt{(1-\wt{M}_0)(1 + 3 \wt{M}_0)}\right) .
\label{eq:wtT-}
\end{equation}

Having solved the equation of motion (\ref{eq:delpot=0}) including
the ghost part, our next task is to consider the BRST invariance
condition (\ref{eq:BRSTinv}). As explained in the introduction, this
condition fixes the coefficients $f_n$ in the BRST operator $\calQ$
(\ref{eq:calQ}) of VSFT rather than gives further constraint on the
solution $\ket{\phic}$.

Using the formula obtained by differentiating (\ref{eq:formulaF}) with
respect to $\zeta_i$, the second term of (\ref{eq:BRSTinv}) is
reduced to the form of $\ket{\phic}$ operated by $c^\dagger_n$:
\begin{align}
&{}_1\bra{\phic}{}_2\bra{\phic}\sum_{r=1}^3\sum_{n\ge 1}
c^{\dagger (r)}_n \W{r3}{n0}\ket{\widehat{V}}_{123}
\Bigr\vert_{p_r=0}
=\calNc \left[\det(1-S\calV)\right]^{-13}
\det(1-\wt{S}\wt{\calV})\nn\\
&\qquad\qquad\qquad\qquad
\times \sum_{n\ge 1}
\left[\wt{\bm{v}}_0 +
(\wt{V}_+,\wt{V}_-)(1-\wt{S}\wt{\calV})^{-1}\wt{S}
\Pmatrix{\wt{\bm{v}}_- \\ \wt{\bm{v}}_+}\right]_n
c^{\dagger (3)}_n \ket{\phic}_3 ,
\label{eq:phiphicWv}
\end{align}
where the vectors $\wt{\bm{v}}_\alpha$ are defined by
\begin{equation}
\left(\wt{\bm{v}}_0\right)_n = \W{r,r}{n0}, \qquad
\left(\wt{\bm{v}}_\pm\right)_n = \W{r,r\pm 1}{n0} .
\label{eq:bmw}
\end{equation}
Then, using (\ref{eq:calNc}) and
\begin{equation}
\calC_n\ket{\phic} = \sum_{m\ge 1}c_m^\dagger
\left(C_{mn} - \wt{S}_{mn}\right)\ket{\phic} ,
\label{eq:calCphi}
\end{equation}
the BRST invariance condition (\ref{eq:BRSTinv}) holds if
$f_n$ ($n\ge 1$) are given by
\begin{align}
\bm{f} &= (C-\wt{S})^{-1}\left[\wt{\bm{v}}_0 +
(\wt{V}_+,\wt{V}_-)(1-\wt{S}\wt{\calV})^{-1}\wt{S}
\Pmatrix{\wt{\bm{v}}_-\\ \wt{\bm{v}}_+}\right]
\nn\\
&= (1-\wt{T})^{-1}\left[\wt{\bm{v}}_0 +
(\wt{M}_+,\wt{M}_-)(1-\wt{T}\wt{\calM})^{-1}\wt{T}
\Pmatrix{\wt{\bm{v}}_+\\ \wt{\bm{v}}_-}\right] ,
\label{eq:f}
\end{align}
where we have used that
\begin{equation}
C\wt{\bm{v}}_0 = \wt{\bm{v}}_0,\quad
C\wt{\bm{v}}_\pm = \wt{\bm{v}}_\mp .
\label{eq:Cwtv}
\end{equation}
The vector $\bm{f}$ obtained this way satisfies $C\bm{f}=\bm{f}$,
namely, $f_{2n+1}=0$.

As explained in \cite{VSFT1}, a homogeneous field redefinition
$\Psi\!\to\!\exp\left(\sum_{n\ge 1}\epsilon_n K_n\right)\Psi$
with $K_n=L_n-(-1)^n L_{-n}$ maps VSFT into another VSFT having
$\calQ$ with different coefficients $\f_n$.
Therefore, $\exp\left(\sum_{n\ge 1}\epsilon_n K_n\right)\Psic$ is a
solution to the equation of motion in VSFT with $f_n$ different
from (\ref{eq:f}).
However, this field redefinition takes our classical solution
$\Psic$ away from the Siegel gauge.

\section{Fluctuation spectrum around $\bm{\Psic}$}

Since we have obtained a classical solution $\Psic$ in VSFT, let us
next examine whether $\Psic$ represents the perturbative open
string vacuum.
Expanding the original string field $\Psi$ in VSFT as
\begin{equation}
\Psi = \Psic + \Phi ,
\label{eq:Psi=Psic+Phi}
\end{equation}
with $\Phi$ being the fluctuation, the VSFT action (\ref{eq:SV}) is
expressed as
\begin{equation}
\SV[\Psi] = \SV[\Psic] -K\left(
\frac12 \Phi\cdot\calQB\Phi + \frac13 \Phi\cdot(\Phi * \Phi)\right) ,
\label{eq:SV=SVc+S}
\end{equation}
where $\calQB$ is defined by
\begin{equation}
\calQB\Phi=\calQ\,\Phi + \Psic *\Phi +\Phi *\Psic .
\label{eq:calQB}
\end{equation}
The new BRST operator $\calQB$ satisfies the nilpotency and the
Leibniz rule on the $*$-product.

What we have to test for confirming that $\Psic$ represents the
perturbative open string vacuum are the following two:
\begin{itemize}
\item
Whether the quadratic term of the fluctuation,
$(1/2)\Psi\!\cdot\!\calQB\Psi$, supports the known perturbative open
string spectrum.

\item
Whether $\SV[\Psic]$ has the expected value of the D25-brane
tension:
\begin{equation}
-\SV[\Psic]= T_{25} V_{26} ,
\label{eq:SVc=T25V26}
\end{equation}
where $V_{26}=\int\!d^{26}x$ is the space-time volume.

\end{itemize}

The test of (\ref{eq:SVc=T25V26}) needs the expression of
$T_{25}$ and hence that of the open string coupling constant $\go$
in terms of the parameters in VSFT.
Since $\go$ is defined by the three-tachyon on-shell amplitude,
what we have to examine first of all is the fluctuation spectrum,
namely, whether $(1/2)\Psi\!\cdot\!\calQB\Psi$ really contains
tachyon and photon etc.

\subsection{Tachyon wave function}

We shall construct the tachyon wave function $\Phit$ which is
a scalar solution to
\begin{equation}
\calQB \Phit=0 ,
\label{eq:calQBPhit=0}
\end{equation}
and carries center-of-mass momentum $p^2=1$ (recall that we are
adopting the convention $\ap=1$).
Of course, $\Phit$ must not be $\calQB$-exact.
Since $\Phit$ should be twist invariant, $\Omega\Phit=\Phit$,
(\ref{eq:calQBPhit=0}) is rewritten as
\begin{equation}
\calQ\,\Phit + (1+\Omega)\left(\Psic *\Phit\right)=0 .
\label{eq:calQPhit+(1+Omega)PsicPhit=0}
\end{equation}
Let us take the Siegel gauge for $\Phit$:
\begin{equation}
\ket{\Phit}=b_0\ket{\phit} .
\label{eq:Phit=b0phit}
\end{equation}
Then, the wave equation (\ref{eq:calQPhit+(1+Omega)PsicPhit=0})
consists of the following two:
\begin{align}
&\ket{\phit}_3
+(1+\Omega_3)
\,{}_1\bra{\phic}{}_2\braket{\phit}{\widehat{V}}_{123}
\Bigr\vert_{p_1=0,\,p_2=-p_3}
=0 ,
\label{eq:Eqphit1}
\\
&\sum_{n \ge 1}\f_n\calC_n\ket{\phit}_3
+(1+\Omega_3)
\, {}_1\bra{\phic}{}_2\bra{\phit}\sum_{r=1}^3\sum_{n\ge 1}
c^{\dagger (r)}_n \W{r3}{n0}\ket{\widehat{V}}_{123}
\Bigr\vert_{p_1=0,\,p_2=-p_3}
=0 .
\label{eq:Eqphit2}
\end{align}
Now, let us try the following $\ket{\phit}$ obtained as a simple
modification on $\ket{\phic}$ (\ref{eq:phic}):
\begin{align}
\ket{\phit}&=\frac{\calNt}{\calNc}
\,\exp\!\left(-\sum_{n\ge 1}t_n a^\dagger_n a_0\right)\ket{\phic}
\nn\\
&=\calNt\exp\left(-\frac12 \sum_{n,m\ge 1}
a^\dagger_n S_{nm}a^\dagger_m + \sum_{n,m\ge 1}
c^\dagger_n \wt{S}_{nm} b^\dagger_m
-\sum_{n\ge 1}t_n a^\dagger_n a_0 \right)\ket{0} .
\label{eq:phit}
\end{align}
In particular, the ghost part of $\ket{\phit}$ is the same as that of
$\ket{\phic}$.
Though not written explicitly, $\ket{\phit}$ carries
non-vanishing momentum in contrast with $\ket{\phic}$ which is
translationally invariant.
Since $\ket{\phit}$ is twist invariant, the coefficient $t_n$
is non-vanishing only for even $n$, namely, the vector $\bm{t}$
satisfies
\begin{equation}
C\bm{t}=\bm{t} .
\label{eq:Ct=t}
\end{equation}
Moreover, $\bm{t}$ is real due to the hermiticity
(\ref{eq:hermiticity}).
The normalization factor $\calNt$ for $\ket{\phit}$ is not determined
by (\ref{eq:Eqphit1}) and (\ref{eq:Eqphit2}) which are linear in
$\ket{\phit}$. It is fixed by a condition given in the next section.

The second term of (\ref{eq:Eqphit1}) for $\ket{\phit}$ of
(\ref{eq:phit}) is calculated by using (\ref{eq:formulaB}) and is
given by
\begin{equation}
{}_1\bra{\phic}{}_2\braket{\phit}{\widehat{V}}_{123}
\Bigr\vert_{p_1=0,\,p_2=-p_3}
= -\frac{\calNt}{\calNc}
\exp\left(-\sum_{n\ge 1}u_n a^{(3)\dagger}_n a^{(3)}_0
-\frac12\,G\,(a^{(3)}_0)^2\right)\ket{\phic}_3 ,
\label{eq:phicphitV}
\end{equation}
where $u_n$ and $G$ on the RHS are
\begin{align}
&\bm{u}=\bm{v}_{0}- \bm{v}_{-} +(V_+,V_-)(1-S\calV)^{-1}S
\Pmatrix{\bm{v}_- - \bm{v}_+ \\ \bm{v}_+ - \bm{v}_0}
+(V_+,V_-)(1-S\calV)^{-1}\Pmatrix{0\\ \bm{t}} ,
\label{eq:u}
\\[2ex]
&G =2\,V_{00} + (\bm{v}_{-}-\bm{v}_{+},
\bm{v}_{+}-\bm{v}_{0})(1-S\calV)^{-1} S
\Pmatrix{\bm{v}_- - \bm{v}_+ \\ \bm{v}_+ - \bm{v}_0}
\nn\\
&\qquad\qquad
+ 2\,(\bm{v}_{-}-\bm{v}_{+},\bm{v}_{+}-\bm{v}_{0})
(1-S\calV)^{-1}\Pmatrix{0\\ \bm{t}}
+(0,\bm{t})\calV (1-S\calV)^{-1}\Pmatrix{0\\ \bm{t}} .
\label{eq:G}
\end{align}
Eq.\ (\ref{eq:phicphitV}) implies that $\ket{\phit}$ of
(\ref{eq:phit}) becomes a solutions to (\ref{eq:Eqphit1})
if $\bm{t}$ satisfies
\begin{equation}
\bm{t}=\bm{u} ,
\label{eq:t=u}
\end{equation}
and the center-of-mass momentum
$a_0=\sqrt{2}\,p$ carried by $\ket{\phit}$ satisfies
\begin{equation}
\exp\left(-\frac12\,G\,(a_0)^2\right)=\frac12 .
\label{eq:p^2fromG}
\end{equation}
Note that the twist operator $\Omega_3$ in (\ref{eq:Eqphit1}) is
effectively equal to the identity since (\ref{eq:phicphitV}) is twist
invariant under (\ref{eq:t=u}).
Moreover, in this case, the second equation (\ref{eq:Eqphit2}) is also
satisfied since the ghost part of $\ket{\phit}$ is the same as that
of $\ket{\phic}$.

Therefore, what we have to do is to first determine the vector $\bm{t}$
satisfying (\ref{eq:t=u}) and then to check whether the momentum $p$
satisfying (\ref{eq:p^2fromG}) really reproduces the tachyon mass,
$p^2=1$.

\subsection{Determination of $\bm{t}$}

Let us solve (\ref{eq:t=u}) for $\bm{t}$, namely,
\begin{align}
&\bm{t}= -\vvpz
+(M_+,M_-)(1-T\calM)^{-1} T
\Pmatrix{\vvpz-\vvmz\\ \vvmz}
+(M_+,M_-)(1-T\calM)^{-1}\Pmatrix{0\\ \bm{t}} ,
\label{eq:Eqfort}
\end{align}
which is obtained by multiplying (\ref{eq:t=u}) by $C$ on the left
and using (\ref{eq:Cv=v}),(\ref{eq:vvpmz}) and (\ref{eq:Ct=t}).
After a tedious calculation using
\begin{equation}
(1-S\calV)^{-1}=\left(1-T\calM\right)^{-1}
=\left[(1-M_0)(1+T)\right]^{-1}
\Pmatrix{1-T M_0 & T M_+\\
           T M_- & 1- T M_0} ,
\label{eq:new(1-TM)^-1}
\end{equation}
valid for $T$ satisfying (\ref{eq:M0T^2-(1+M0)T+M0=0})
(recall (\ref{eq:(1-TM)^-1})),
and the formulas (\ref{eq:M+M+M=1})--(\ref{eq:M^2-M=MM}),
(\ref{eq:MvI}) and (\ref{eq:MvII}) for $M_\alpha$ and $\bm{v}_\alpha$,
eq.\ (\ref{eq:Eqfort}) multiplied by $(1-M_0)(1+T)$ is reduced to the
following simple equation:
\begin{equation}
(M_{+}+M_{-}T)\bm{t}=(T-1)\left(\vvpz - T\vvmz\right) .
\label{eq:simpleeqT}
\end{equation}

Note first that we cannot invert $M_+ + M_- T$ to solve
(\ref{eq:simpleeqT}) since the following identity holds for $T$
satisfying (\ref{eq:M0T^2-(1+M0)T+M0=0}):\footnote{
For consistency, the RHS of (\ref{eq:simpleeqT}) operated by
$M_{-}+M_{+}T$ must vanish. This is indeed the case since we have
$(M_{+}+M_{-}T)\left(\vvpz - T\vvmz\right)=0$.
}
\begin{equation}
(M_{-} + M_{+}T)(M_{+}+M_{-}T)=0 .
\label{eq:(M+MT))M+MT)=0}
\end{equation}
Therefore, solution to (\ref{eq:simpleeqT}) is not unique.
However, twist invariant solution satisfying (\ref{eq:Ct=t}) does
exist uniquely as we shall show.
A twist invariant solution $\bm{t}$ must satisfy both
(\ref{eq:simpleeqT}) and the same one multiplied by $C$:
\begin{equation}
(M_{-}+M_{+}T)\bm{t}=(T-1)\left(\vvmz - T\vvpz\right) .
\label{eq:CsimpleeqT}
\end{equation}
As the solution of the equation obtained by adding
(\ref{eq:simpleeqT}) and (\ref{eq:CsimpleeqT}),
we get
\begin{equation}
\bm{t}=-\left[(1-M_0)(1+T)\right]^{-1}(1-T)^2
\left(\vvpz + \vvmz\right) .
\label{eq:t}
\end{equation}
It is a straightforward exercise to show that the solution
(\ref{eq:t}) also satisfies the equation obtained by subtracting
the two.

\subsection{Tachyon mass}

Having obtained the vector $\bm{t}$, we shall proceed to the
determination of the mass of the state $\Phit$.
We shall first calculate $G$ (\ref{eq:G}),
which is rewritten as
\begin{align}
&G =2\,V_{00} + (\vvmz-\vvpz,\vvpz)(1-T\calM)^{-1} T
\Pmatrix{\vvpz-\vvmz \\ \vvmz}
\nn\\
&\qquad\qquad
+ 2\,(\vvmz-\vvpz,\vvpz)
(1-T\calM)^{-1}\Pmatrix{0\\ \bm{t}}
+(0,\bm{t})\calM (1-T\calM)^{-1}\Pmatrix{0\\ \bm{t}} .
\label{eq:Gagain}
\end{align}
and then obtain the mass from (\ref{eq:p^2fromG}).
We present here a concise expression of $G$ with $\bm{t}$ given by
(\ref{eq:t}):
\begin{equation}
G =\vvpz G_{++}\vvpz + \vvpz G_{+-}\vvmz ,
\label{eq:G=vGv+vGv}
\end{equation}
with
\begin{align}
G_{++} &=
-2\,M_0\,(1-M_0)^{-1}(1+3 M_0)^{-2}
\Bigl((3 M_0-1)T + 9 M_0^2- 3\Bigr) ,
\label{eq:G++}
\\
G_{+-} &=
(1+M_0)\,(1-M_0)^{-1}(1+3 M_0)^{-2}
\Bigl((3 M_0-1)T + 9 M_0^2- 3\Bigr) .
\label{eq:G+-}
\end{align}
The $2\,V_{00}$ term in (\ref{eq:Gagain}) has been included in
(\ref{eq:G=vGv+vGv}) by use of (\ref{eq:vv=V00}).
Derivation of (\ref{eq:G=vGv+vGv}) is rather tedious and is
summarized in appendix \ref{sec:G}.

\begin{table}[htbp]
\begin{center}
\begin{tabular}[b]{|r|l|l|}
\hline
$L$~ & ~~~$G$ & ~~$-\mt^2$\\
\hline\hline
$10$ & $0.6493$ & $1.06746$ \\
$50$ & $0.6732$ & $1.02958$ \\
$100$ & $0.6773$ & $1.02333$ \\
$150$ & $0.6791$ & $1.02072$ \\
$200$ & $0.6801$ & $1.01918$ \\
$250$ & $0.6808$ & $1.01812$ \\
$300$ & $0.6813$ & $1.01734$ \\
\hline
\end{tabular}
\hspace{5ex}
\begin{tabular}[b]{|r|l|l|}
\hline
$L$~ & ~~~$G$ & ~~$-\mt^2$\\
\hline\hline
$9$ & $0.6665$ & $1.04000$ \\
$49$ & $0.6760$ & $1.02534$ \\
$99$ & $0.6786$ & $1.02141$ \\
$149$ & $0.6799$ & $1.01951$ \\
$199$ & $0.6807$ & $1.01830$ \\
$249$ & $0.6813$ & $1.01744$ \\
$299$ & $0.6817$ & $1.01678$ \\
\hline
\end{tabular}
\caption{
$G$ and $-\mt^2$ for various values of the cutoff $L$.
The left (right) table shows the result for even (odd) $L$.
}
\label{tab:G}
\end{center}
\end{table}

We have not succeeded in analytically evaluating the value of $G$.
Instead, we have calculated $G$ numerically by level truncation
approximation. Namely, we restrict the indices $n$, $m$ of
the matrix $M_0$ and the vectors $\bm{v}_{\pm 0}$ to
$1\le n,m \le L$ and calculate $G$ (\ref{eq:G=vGv+vGv}) for various
values of the cutoff $L$.
The matrix $T$ for a finite $L$ is defined by (\ref{eq:T-}).
Table \ref{tab:G} summarizes the result of our calculation.
For each value of $L$, we calculated $G$ and the corresponding mass
squared $\mt^2$ obtained from (\ref{eq:p^2fromG}):
\begin{equation}
e^{G\mt^2}=\frac12 .
\label{eq:e^Gmt^2=1/2}
\end{equation}
As seen from the table, $\mt^2$ tends to the expected value of the
tachyon mass squared, $-1$, as we increase the cutoff $L$
(a slightly better result is obtained for odd $L$ than even $L$).
The value of $-\mt^2$ at $L=\infty$ predicted by a fitting
function of the form $\sum_{k=0}^3 c_k (\ln L)^{-k}$ is
$1.0013$ ($0.9947$) for even (odd) $L$.
Though rigorous and analytic evaluation of $\mt^2$ is of course
desired, our analysis here strongly supports that the wave function
$\Phit$ really represents the tachyon mode on the perturbative open
string vacuum.

\subsection{Massless vector mode}
\label{sec:phip}

In the previous subsections, we have succeeded in identifying the
tachyon mode as a fluctuation mode around the VSFT solution $\Psic$.
As another test of the fluctuation spectrum around $\Psic$,
let us try constructing the wave function $\Phip$ representing the
massless vector state on the perturbative vacuum.
Here again, we try the Siegel gauge solution
\begin{equation}
\ket{\Phip}=b_0\ket{\phip} ,
\end{equation}
and consider $\ket{\phip}$ of the following form:
\begin{equation}
\ket{\phip}=\left(
\sum_{n=1,3,5,\ldots}d^\mu_n a^{\mu\dagger}_n\right)\ket{\phit} .
\label{eq:phip}
\end{equation}
Since we are interested in the massless states which should be odd
under the twist transformation, we assume that vector $d_n^\mu$
has nonvanishing (and real) components only for odd $n$:
\begin{equation}
C\bm{d}^\mu=-\bm{d}^\mu .
\label{eq:Cd=-d}
\end{equation}
Then, the wave equation $\calQB\Phip=0$ consists of the following two
for $\phip$:
\begin{align}
&\ket{\phip}_3
+ \left(1-\Omega_3\right)
\,{}_1\bra{\phic}{}_2\braket{\phip}{\widehat{V}}_{123}
\Bigr\vert_{p_1=0,\, p_2=-p_3}
=0 ,
\label{eq:Eqphip1}
\\
&\sum_{n \ge 1}\f_n\calC_n\ket{\phip}_3
+\left(1-\Omega_3\right)
\, {}_1\bra{\phic}{}_2\bra{\phip}\sum_{r=1}^3\sum_{n\ge 1}
c^{\dagger (r)}_n \W{r3}{n0}\ket{\widehat{V}}_{123}
\Bigr\vert_{p_1=0,\, p_2=-p_3}=0 .
\label{eq:Eqphip2}
\end{align}

Let us consider the second term of (\ref{eq:Eqphip1}).
Using the formula (\ref{eq:formulaB}) differentiated with respect to
$K_i$, we have
\begin{align}
&{}_1\bra{\phic}{}_2\braket{\phip}{\widehat{V}}_{123}
=-\exp\left(-\frac12 G(a^{(3)}_0)^2\right)
\Biggl\{
-\sum_{n\ge 1}
\left[(V_+,V_-)(1-S\calV)^{-1}\Pmatrix{0\\ \bm{d}^\mu}\right]_n
a^{(3)\mu\dagger}_{n}
\nn\\
&\qquad\qquad
-\Bigl[
(0,\bm{t})\calV + (\bm{v}_- -\bm{v}_+,\bm{v}_+ -\bm{v}_0)\Bigr]
(1-S\calV)^{-1}\Pmatrix{0\\ \bm{d}^\mu}a^{(3)\mu}_0
\Biggr\}\ket{\phit}_3 .
\label{eq:phicphipV}
\end{align}
The $a_0^{(3)\mu}$ term in (\ref{eq:phicphipV}) does not survive the
twist-odd projection.
For the $a^{(3)\mu\dagger}_{n}$ term operated by $1-\Omega_3$, we use
the following identity valid for an arbitrary $\bm{d}^\mu$ and proved
by using (\ref{eq:new(1-TM)^-1}), (\ref{eq:M^2-M=MM}) and
(\ref{eq:Cd=-d}):
\begin{equation}
(1-C)(V_+,V_-)(1-S\calV)^{-1} \Pmatrix{0\\ \bm{d}^\mu}
= -\bm{d}^\mu .
\label{eq:identityd}
\end{equation}
Therefore, we have
\begin{equation}
(1-\Omega_3){}_1\bra{\phic}{}_2\braket{\phip}{\widehat{V}}_{123}
=-\exp\left(-\frac12 G(a^{(3)}_0)^2\right)\ket{\phip}_3 ,
\label{eq:(1-Omega)phicphipV}
\end{equation}
for any $\bm{d}^\mu$ parameterizing the state $\ket{\phip}$.
Eq.\ (\ref{eq:(1-Omega)phicphipV}) implies that the first of the wave
equation, (\ref{eq:Eqphip1}), holds if the center-of-mass momentum
carried by $\ket{\phip}$ satisfies $p^2=0$. Namely, our $\ket{\Phip}$
represents a massless state for any $\bm{d}^\mu$.

Once the first equation (\ref{eq:Eqphip1}) is satisfied, the second
one (\ref{eq:Eqphip2}) holds automatically since the ghost part of
${}_1\bra{\phic}{}_2\bra{\phip}\sum_{r=1}^3\sum_{n\ge 1}
c^{\dagger (r)}_n \W{r3}{n0}\ket{\widehat{V}}_{123}$
is the same as that for $\phit$ and is twist invariant.
Therefore, the whole wave equation $\calQB\Phip=0$ is satisfied by the
present $\Phip$ only if we set $p^2=0$.
In particular, the vector $\bm{d}^\mu$  can be completely arbitrary.
However, this is an unwelcome fact.
We expect that the wave equation $\calQB\Phip=0$ gives the
transversality condition on the polarization vector as well as the
on-shell condition as in the case of ordinary BRST operator
$\QB=\sum_n c_{-n}\left(L^{\rm matt}_n + (1/2)L^{\rm gh}_n\right)$.
In the present case, $\bm{d}^\mu$ is fixed neither as a space-time
vector nor as a vector in the level number space.

We do not know whether this trouble is merely an artifact of taking
the Siegel gauge for $\Phip$ or it indicates that our classical
solution $\Psic$ does  not represent the perturbative open string
vacuum. A more detailed analysis of the massless fluctuation mode
space including the ghost sector is necessary to get a definite
answer. Here, as a partial support for our identification of
$\ket{\Phip}$ as the massless vector state, we present the result of
our calculation of the tachyon-tachyon-vector coupling
$\bbbk{\phit}{\phit}{\phip}{\widehat{V}}$ using the present
$\ket{\phip}$ of (\ref{eq:phip}).
We find that its momentum dependence takes a familiar form
$(p_1-p_2)_\mu\zeta^\mu$ with the ``effective polarization vector''
$\zeta^\mu$ given by\footnote{
The original expression using the quantities defined in sec.\ 5 is
$(A_0^\mu\bm{t}\calV_3 -\bm{V}^\mu)(1-S\calV_3)^{-1}
(0,0,\bm{d}^\mu)^\T$. The derivation of (\ref{eq:effectivepolvec}) is
similar to that of $H$ (\ref{eq:H=vHv+vHv}) explained in appendix
\ref{sec:H}.
}
\begin{equation}
\zeta^\mu=\vvpz\!\left[(1-M_0)(1+3 M_0)^2\right]^{-1}
M_0^2 (2-T)\,\bm{d}^\mu .
\label{eq:effectivepolvec}
\end{equation}

\section{Potential height}

In the previous section, we found that the fluctuation around $\Psic$
contains at least the tachyon and the massless vector modes.
In this section, we shall carry out the test of (\ref{eq:SVc=T25V26})
concerning the energy density of the solution $\Psic$.
For this purpose, we shall first calculate various quantities:
energy density $-\SV[\Psic]/V_{26}$, normalization factor $\calNt$
of the tachyon wave function, and the open string coupling constant
$\go$.
After these preparations, we shall proceed to the examination of
(\ref{eq:SVc=T25V26}).

\subsection{Energy density $\bm{\calEc}$}

In this subsection, we shall evaluate the energy density $\calEc$
of the solution $\Psic$.
Using the equation of motion for $\Psic$,
$\calQ \Psic + \Psic\! *\!\Psic =0$, we have
\begin{align}
\SV[\Psic]
&=-\frac{K}{6}\intund{b_0,x}\bra{\Psic}\calQ\ket{\Psic}
=-\frac{K}{6}\braket{\phic}{\phic}\,V_{26}
\nn\\
&=-\frac{K}{6}\,\calNc^2\left[\det(1-S^2)\right]^{-13}
\det(1-\wt{S}^2)\,V_{26} .
\label{eq:SV[Psic]}
\end{align}
When $T$ satisfies (\ref{eq:M0T^2-(1+M0)T+M0=0}),
the normalization factor $\calNc$ (\ref{eq:calNc}) is written as
\begin{equation}
\calNc= -\left(
\det\left[(1-M_0)(1+T)\right]\right)^{13}
\left(\det\left[(1-\wt{M}_0)(1+\wt{T})\right]\right)^{-1} .
\label{eq:calNyet}
\end{equation}
Then, using (\ref{eq:T-}) and (\ref{eq:wtT-}), in particular,
\begin{equation}
\frac{1+T}{1-T}=\sqrt{\frac{1+3 M_0}{1-M_0}} ,
\end{equation}
the energy density of our solution $\Psic$ is given by
\begin{align}
\calEc=-\frac{\SV[\Psic]}{V_{26}}=
\frac{K}{6}
&\left(\det\left[(1-M_0)^{3/2}(1+3 M_0)^{1/2}\right]
\right)^{13}
\nn\\
&\qquad\times
\left(\det\left[(1-\wt{M}_0)^{3/2}(1+3 \wt{M}_0)^{1/2}\right]
\right)^{-1} .
\label{eq:calEc}
\end{align}

\subsection{Normalization factor $\bm{\calNt}$}

We determine the normalization factor $\calNt$ in (\ref{eq:phit})
by the requirement that the ``tachyon field'' $\varphit(x)$
which appear in the expansion of the fluctuation $\Phi$
(\ref{eq:Psi=Psic+Phi}) as
\begin{equation}
\ket{\Phi}=\ket{\Phit}\varphit(x) + \cdots ,
\label{eq:expandPsi}
\end{equation}
has canonical kinetic term in $-K(1/2)\Phi\cdot\calQB\Phi$
(see eq.\ (\ref{eq:SV=SVc+S})).
Therefore, let us consider the following quantity for $\Phit$ with
off-shell $p^2$ ($p^2\ne -\mt^2$)
(recall eqs.\ (\ref{eq:Eqphit1}) and (\ref{eq:phicphitV})):
\begin{align}
&\int\!db_0\,\bra{\Phit}\calQB\ket{\Phit}
={}_3\bra{\phit}\left(\ket{\phit}_3
+ 2\,{}_1\bra{\phic}{}_2\braket{\phit}{\widehat{V}}_{123}\right)
\nn\\
&=\left(1 -2\,e^{-G p^2}\right)
\braket{\phit}{\phit}
=\left(1 -e^{-G(p^2+\mt^2)}\right)
\braket{\phit}{\phit} ,
\label{eq:PhitcalQBPhit}
\end{align}
where we have used (\ref{eq:e^Gmt^2=1/2}). The inner product
$\braket{\phit}{\phit}$ is given by
\begin{equation}
\braket{\phit}{\phit}
=\calNt^2 \left[\det(1-S^2)\right]^{-13}\det(1-\wt{S}^2)
\,\exp\!\left(2\,\bm{t}\,(1+T)^{-1}\bm{t}\,p^2\right) .
\end{equation}
Taylor expanding (\ref{eq:PhitcalQBPhit}) around $p^2=-\mt^2$,
we have
\begin{align}
K\!\int\!db_0\,\bra{\Phit}\calQB\ket{\Phit}
\underset{p^2\sim -\mt^2}{\sim}
&K\calNt^2 G \left[\det(1-S^2)\right]^{-13}\det(1-\wt{S}^2)
\nn\\
&\qquad\times
\exp\!\left(-2\,\bm{t}\,(1+T)^{-1}\bm{t}\,\mt^2\right)
\cdot\left(p^2 + \mt^2\right) ,
\end{align}
from which $\calNt$ is read off as
\begin{align}
\calNt=\left(\frac{1}{K G}\right)^{1/2}
\left[\det(1-S^2)\right]^{13/2}\left[\det(1-\wt{S}^2)\right]^{-1/2}
\exp\Bigl(\bm{t}\,(1+T)^{-1}\bm{t}\,\mt^2\Bigr) .
\label{eq:calNt}
\end{align}

\subsection{$\bm{\go}$ as three-tachyon on-shell coupling}

The open string coupling constant $\go$ is defined to be the
three-tachyon on-shell amplitude.
Using the tachyon wave function $\Phit$ with normalized kinetic term,
$\go$ is given by
\begin{align}
\go &=K\,\bbbk{\phit}{\phit}{\phit}{\widehat{V}}
\Bigr\vert_{p_1^2=p_2^2=p_3^2=-\mt^2}
\nn\\
&=K\calNt^3 {}_{123}\bra{0}\exp\left(
-\frac12 \bm{A}S\bm{A} - A_0\bm{t}\bm{A}
-\sum_{r=1}^3\sum_{n,m\ge 1}c^{(r)}_n \wt{S}_{nm} b^{(r)}_m
\right)
\nn\\
&\qquad\qquad\times
\exp\left(
-\frac12\,\bm{A}^\dagger\calV_3\bm{A}^\dagger
-\bm{A}^\dagger\bm{V} -\frac12 V_{00}(A_0)^2
+ \sum_{r,s=1}^3\sum_{n,m\ge 1}
c^{(r)\dagger}_n \W{rs}{nm} b^{(s)\dagger}_m
\right)\ket{0}_{123}
\nn\\
&=K\calNt^3\,\left[\det(1-S\calV_3)\right]^{-13}
\det(1-\wt{S}\wt{\calV}_3)
\,\exp\Biggl\{
-\frac12\bm{V}(1-S\calV_3)^{-1}S\bm{V}
\nn\\
&\qquad\qquad\qquad
+\bm{V}(1-S\calV_3)^{-1}\bm{t} A_0
-\frac12 A_0\bm{t}\calV_3(1-S\calV_3)^{-1}\bm{t} A_0
-\frac12 V_{00}(A_0)^2
\Biggr\} ,
\label{eq:KphitphitphitV}
\end{align}
with various new quantities defined by
\begin{align}
&\bm{A}=\Pmatrix{a^{(1)}_n\\a^{(2)}_n\\a^{(3)}_n} ,
\quad
A_0=\Pmatrix{a^{(1)}_0\\a^{(2)}_0\\a^{(3)}_0} ,
\quad
\calV_3 =\Pmatrix{V_0&V_+&V_-\\ V_-&V_0&V_+\\ V_+&V_-&V_0} ,
\quad
\wt{\calV}_3 =\Pmatrix{\wt{V}_0&\wt{V}_+&\wt{V}_-\\
\wt{V}_-&\wt{V}_0&\wt{V}_+\\ \wt{V}_+&\wt{V}_-&\wt{V}_0} ,
\nn\\[2ex]
&\bm{V} =
\Pmatrix{\bm{v}_0&\bm{v}_+&\bm{v}_-\\
         \bm{v}_-&\bm{v}_0&\bm{v}_+\\
         \bm{v}_+&\bm{v}_-&\bm{v}_0}\!\!
\Pmatrix{a^{(1)}_0\\a^{(2)}_0\\a^{(3)}_0}
=\Pmatrix{a^{(2)}_0\vvpz+a^{(3)}_0\vvmz\\
          a^{(3)}_0\vvpz+a^{(1)}_0\vvmz\\
          a^{(1)}_0\vvpz+a^{(2)}_0\vvmz}
=\Pmatrix{a^{(2)}_0& a^{(3)}_0\\
          a^{(3)}_0& a^{(1)}_0\\
          a^{(1)}_0& a^{(2)}_0}\Pmatrix{\vvpz\\ \vvmz} .
\label{eq:newquantities}
\end{align}
The boldface letters are vectors with respect to the level number $n$.
Though we have omitted the transpose symbol for the vectors in
(\ref{eq:KphitphitphitV}), how they form inner products should be
evident.

Each term in the exponent of (\ref{eq:KphitphitphitV}) is reduced to
$(A_0)^2=\sum_{r=1}^3 (a^{(r)}_0)^2$ times a quantity consisting of the
Neumann coefficients
(their calculations are summarized in appendix \ref{sec:H}).
Using the on-shell condition, $(A_0)^2=-6\mt^2$, and substituting
$\calNt$ of (\ref{eq:calNt}), the square of $\go$ is given by
\begin{align}
\go^2 &=\frac{1}{K G^3}\Bigl(
\det(1-T^2)^{-3} \det(1-S\calV_3)^2\Bigr)^{-13}
\nn\\
&\qquad\qquad\times
\det(1-\wt{T}^2)^{-3} \det(1-\wt{S}\wt{\calV}_3)^2\,
\exp\left(6 H \mt^2\right) ,
\label{eq:go^2}
\end{align}
where $H$ has the following expression similar to $G$
(\ref{eq:G=vGv+vGv}):
\begin{equation}
H=\vvpz H_{++}\vvpz + \vvpz H_{+-}\vvmz ,
\label{eq:H=vHv+vHv}
\end{equation}
with
\begin{align}
&H_{++}=3\,M_0 \left[(1-M_0)(1+3 M_0)^2\right]^{-1}
\left(1+4 M_0-3 M_0^2 +2\,M_0 T \right) ,
\nn\\
&H_{+-}=-\frac32\,(1+M_0)
\left[(1-M_0)(1+3 M_0)^2\right]^{-1}
\left(1+4 M_0-3 M_0^2 +2\,M_0 T \right) .
\label{eq:H++H+-}
\end{align}

Note that the three-tachyon coupling calculated here does not take a
form which vanishes when the external momenta are put on-shell.
This implies that the tachyon wave function $\Phit$ constructed in
sec.\ 4 is not $\calQB$-exact.

\subsection{$\bm{\calEc/T_{25}}$}
\label{sec:calEc/T25}

Now we are ready to consider the ratio $\calEc/T_{25}$.
The energy density $\calEc$ is given by (\ref{eq:calEc}) and
the D25-brane tension by $T_{25}=1/(2\pi^2\go^2)$
with $\go^2$ (\ref{eq:go^2}).
First, we see that the determinant factors cancel between the two
since we have
\begin{equation}
\det(1-T^2)^{-3}\det(1-S\calV_3)^2
=\det\left[(1-M_0)^{3/2}(1+3 M_0)^{1/2}\right] ,
\label{eq:surprise}
\end{equation}
and the corresponding one for the tilded matrices.
To prove this, first note the following expression for the
$3\times 3$ part of the determinant $\det(1-S\calV_3)$:
\begin{align}
\underset{3\times 3}{\det}(1-S\calV_3)
&=1- 3 M_0 T
+3\left(M_0^2-M_+ M_-\right)T^2
-\left(M_+^3 +M_-^3 + M_0^3 - 3 M_0 M_+ M_-\right)T^3
\nn\\
&= 1 - 3\,M_0 + 3\,T - T^3 ,
\label{eq:det3x3}
\end{align}
where we have used (\ref{eq:M+M+M=1})--(\ref{eq:M^2+M^2+M^2=1})
and (\ref{eq:M0T^2-(1+M0)T+M0=0}) in obtaining the last expression.
Eq.\ (\ref{eq:surprise}) is a consequence of (\ref{eq:det3x3}) and an
identity for $T$ of (\ref{eq:T-}):
\begin{align}
(1-T^2)^{-3}\left(1-3 M_0+3 T- T^3\right)^2
=(1-M_0)^{3/2}(1+3 M_0)^{1/2} .
\end{align}

Therefore, the ratio between the energy density $\calEc$ and the
D25-brane tension $T_{25}$ reads
\begin{equation}
\frac{\calEc}{T_{25}}=
\frac{\pi^2}{3 G^3}\exp\left(6\,\mt^2\,H\right) .
\label{eq:calEc/T25}
\end{equation}
This ratio is given in terms of $G$ (\ref{eq:G=vGv+vGv}), the tachyon
mass squared $\mt^2$ determined by (\ref{eq:e^Gmt^2=1/2}), and $H$
(\ref{eq:H=vHv+vHv}).
The constant $K$ multiplying the VSFT action (\ref{eq:SV}) has been
cancelled out in (\ref{eq:calEc/T25}) as of course it should be.
If we admit that the tachyon mass is correctly reproduced,
namely, $\mt^2=-1$ and $G=\ln 2$
(otherwise the present analysis does not make sense),
(\ref{eq:calEc/T25}) is rewritten as
\begin{equation}
\frac{\calEc}{T_{25}}=
\frac{\pi^2}{3\,(\ln 2)^3}\,e^{-6\,H} .
\label{eq:calEc/T25ln2}
\end{equation}
If our classical solution $\Psic$ in VSFT represents the perturbative
open string vacuum, we should have $\calEc/T_{25}=1$. Namely, the
desired value of $H$ is
$H=(1/6)\ln(\pi^2/(3\ln^3 2))\simeq 0.3817$.

\begin{table}[htbp]
\begin{center}
\begin{tabular}[b]{|r|l|l|}
\hline
$L$~ & ~~~~$H$ & $\calEc/T_{25}$\\
\hline\hline
$10$ & $-0.0418$ & $12.69$ \\
$50$ & $-0.1160$ & $19.81$ \\
$100$ & $-0.1386$ & $22.69$ \\
$150$ & $-0.1500$ & $24.29$ \\
$200$ & $-0.1573$ & $25.39$ \\
$250$ & $-0.1627$ & $26.22$ \\
$300$ & $-0.1668$ & $26.88$ \\
\hline
\end{tabular}
\caption{
$H$ (\ref{eq:H=vHv+vHv}) and $\calEc/T_{25}$ (\ref{eq:calEc/T25ln2})
for various values of the cutoff $L$.
}
\label{tab:H}
\end{center}
\end{table}

We have carried out numerical analysis of $H$ (\ref{eq:H=vHv+vHv})
by cutting off the size of the matrices to $L\times L$.
The result of our calculation given in table \ref{tab:H} is very
different from our expectation.
As the cutoff $L$ is increased, $\calEc/T_{25}$ becomes larger far
beyond the desired value of one.

The most naive and hasty interpretation of this unwanted
behavior of $\calEc/T_{25}$ is that our solution $\Psic$ of VSFT does
not correspond to the perturbative open string vacuum.
However, in the next subsection we shall argue that there are in fact
ambiguities in the calculation of $\calEc/T_{25}$.

\subsection{Reexamination of $\bm{\calEc/T_{25}}$}
\label{sec:subtle}

Recall that, though each of $\calEc$ and $T_{25}$ contains
determinant factors, they are cancelled out in the ratio
(\ref{eq:calEc/T25}).
However, numerical analysis shows that these determinant factors
themselves are not finite quantities, and this fact suggests that
there might be subtle points in the treatment of the determinants
in $\calEc/T_{25}$.
Therefore, we shall reexamine the ratio $\calEc/T_{25}$ by adopting
different treatments than those in the previous subsections:
We do not use the equation of motion for $\Psic$,
$\calQ\Psic+\Psic\! *\!\Psic=0$, in calculating $\calEc$, nor
do we use the commutativity among $M_\alpha$ and various identities
originating from (\ref{eq:M+M+M=1}) and (\ref{eq:MM=M^2-M})
to simplify the determinants

Without using the equation of motion, the action of our
classical solution $\SV[\Psic]$ is given instead of
(\ref{eq:SV[Psic]}) by
\begin{align}
\SV[\Psic]=-K\Biggl(
\frac12\,\calNc^2 &\left[\det(1-S^2)\right]^{-13}\det(1-\wt{S}^2)
\nn\\
&+\frac13\,\calNc^3 \left[\det(1-S\calV_3)\right]^{-13}
\det(1-\wt{S}\wt{\calV}_3)\Biggr)V_{26} .
\end{align}
Then, using  (\ref{eq:calNc}) for $\calNc$ and (\ref{eq:go^2}) for
$\go^2$, the ratio $\calEc/T_{25}$ is given instead of
(\ref{eq:calEc/T25}) by
\begin{align}
\frac{\calEc}{T_{25}}
&=\frac{\pi^2}{3 G^3}\,e^{6\mt^2 H}\, Z ,
\end{align}
where the extra factor $Z$ is
\begin{equation}
Z=3 \left(\frac{\left(\Zmatt\right)^{13}}{\Zgh}\right)^2
- 2 \left(\frac{\left(\Zmatt\right)^{13}}{\Zgh}\right)^3 ,
\label{eq:Z}
\end{equation}
with
\begin{align}
\Zmatt &=\frac{\det(1-S^2)\det(1-S\calV)}{\det(1-S\calV_3)} ,
\label{eq:Zmatt}
\\
\Zgh &=\frac{\det(1-\wt{S}^2)\det(1-\wt{S}\wt{\calV})}
{\det(1-\wt{S}\wt{\calV}_3)} .
\label{eq:Zgh}
\end{align}
If we use the commutativity among $M_\alpha$ and their identities
given in sec.\ 2, we have $\Zmatt=\Zgh=1$ and the previous
expression (\ref{eq:calEc/T25}) is recovered.

\begin{table}[htbp]
\begin{center}
\begin{tabular}[b]{|r|l|l|r|}
\hline
$L$~&~~$\Zmatt$&~~$\Zgh$&$Z$~~~~~\\
\hline\hline
$2$ & $1.0195$ & $1.0487$ & $0.8253$ \\
$4$ & $1.0318$ & $1.0734$ & $0.3912$ \\
$6$ & $1.0396$ & $1.0881$ & $-0.1000$ \\
$8$ & $1.0450$ & $1.0984$ & $-0.5903$ \\
$10$ & $1.0492$ & $1.1062$  & $-1.0644$ \\
$20$ & $1.0615$ & $1.1291$  & $-3.1475$ \\
$30$ & $1.0683$ & $1.1416$  & $-4.8626$ \\
\hline
\end{tabular}
\caption{
$\Zmatt$ (\ref{eq:Zmatt}), $\Zgh$ (\ref{eq:Zgh}) and $Z$ (\ref{eq:Z})
for various values of the cutoff $L$.
}
\label{tab:Z}
\end{center}
\end{table}

Table \ref{tab:Z} shows the result of our numerical calculation of
$\Zmatt$, $\Zgh$ and $Z$ using level truncation. We did not use
the commutativity to reduce $2L\times 2L$ determinant $\det(1-S\calV)$
and $3L\times 3L$ one $\det(1-S\calV_3)$ to $L\times L$ determinants
as we did before. We treated them as they stand (note that both the
commutativity and the non-linear identity (\ref{eq:MM=M^2-M}) are
violated for a finite cutoff $L$). We used (\ref{eq:T-}) as the matrix
$S=CT$ in the determinants.
As seen from table \ref{tab:Z}, both $\Zmatt$ and $\Zgh$ gradually
deviate from one rather than approach it as the cutoff $L$ is
increased. The total factor $Z$ appearing in the ratio $\calEc/T_{25}$
deviates from one much faster due mainly to the large power
$(\Zmatt)^{39}$ in the last term of (\ref{eq:Z}), and it even becomes
negative for larger $L$.

This result itself does not directly remedy the unwanted
behavior of $\calEc/T_{25}$ given in table \ref{tab:H} but rather
worsens it.
We even do not know which of the two, $Z=1$ as in
sec.\ \ref{sec:calEc/T25} or a non-trivial $Z$ here, is a ``correct''
one.
A similar problem may exist also in $H$.
In any case, a lesson we learn from the present analysis is that the
ratio  $\calEc/T_{25}$ is a rather ambiguous quantity.
To give the final answer to the potential height problem for our
classical solution, we have to understand a basic principle which
fixes the ambiguities.

\section{Conclusion}

Our analysis for the VSFT solution $\Psic$ given in this paper is
still incomplete for drawing a conclusion on whether $\Psic$ represents
the perturbative open string vacuum.
Affirmative results are the tachyon and the vector masses (the former
was calculated numerically to give an almost expected value, while the
latter is exactly zero), and the tachyon-tachyon-vector coupling.
However, we got a disappointing result for the potential height
$\calEc/T_{25}$.

Our remaining problems are now evident.
First, we have to clarify the full structure of the fluctuation modes
satisfying the wave equation $\calQB\Phi=0$ besides tachyon and
massless vector modes. It is also necessary to resolve the problem
in constructing the massless vector wave function described in sec.\
\ref{sec:phip}.
The most important and probably a difficult problem is a
reconsideration of the potential height.
As discussed in sec.\ \ref{sec:subtle}, there seems to be ambiguities
in the analysis of $\calEc/T_{25}$, which we have to
settle for giving the final answer.
Although our VSFT classical solution $\Psic$ is given in the Siegel
gauge, it is of interest to consider solutions in other gauges.
There might be gauges where various quantities appearing in
$\calEc/T_{25}$ become less singular.

\section*{Acknowledgments}
We would like to thank I.~Kishimoto, K.~Ohmori, K.~Okuyama,
S.~Shinohara and S.~Teraguchi for valuable discussions and useful
comments.
The work of H.\,H.\ was supported in part by a Grant-in-Aid for
Scientific Research from Ministry of Education, Culture, Sports,
Science, and Technology (\#12640264). The work of T.\,K.\ was
supported in part by a Grant-in-Aid for Scientific Research in a
Priority Area: ``Supersymmetry and Unified Theory of Elementary
Particles''(\#707), from the Ministry of Education, Culture, Sports,
Science, and Technology.

\appendix
\vspace{1.5cm}
\centerline{\Large\bf Appendix}
\appendix

\section{Useful formulas}
\label{sec:formulas}

We quote two basic formulas frequently used in the text for
calculating inner products between squeezed states.
For any bosonic (fermionic) oscillators satisfying the commutation
(anti-commutation) relations,
\begin{equation}
[a_i,a^\dagger_j]=\delta_{ij},
\qquad
\{c_i,b^\dagger_j\}=\{b_i,c^\dagger_j\}=\delta_{ij} ,
\label{eq:commrelation}
\end{equation}
and for the Fock vacuum $\ket{0}$ annihilated by $(a_i,b_i,c_i)$ for
any $i$, we have
\begin{align}
&\bra{0}\exp\left(-\frac12 a_i A_{ij} a_j - K_i a_i\right)
\exp\left(-\frac12 a^\dagger_i B_{ij} a^\dagger_j - J_i a^\dagger _i
\right)\!\ket{0}\nn\\
&\qquad\qquad\qquad
=\left[\det\left(1 - AB\right)\right]^{-1/2}
\exp\left(-\frac12 J P A J -\frac12 K B P K + J P K \right) ,
\label{eq:formulaB}
\\[2ex]
&\bra{0}\exp\left(-c_i F_{ij} b_j\right)
\exp\left(c^\dagger_i G_{ij} b^\dagger_j
+\eta_i b^\dagger_i + c^\dagger_i \zeta_i\right)\!\ket{0}
\nn\\
&\qquad\qquad\qquad
=\det\left(1 - FG\right)
\exp\Bigl(\eta (1 - FG)^{-1} F \zeta\Bigr) ,
\label{eq:formulaF}
\end{align}
where $P$ in (\ref{eq:formulaB}) is $P=(1-AB)^{-1}$, and $\eta_i$ and
$\zeta_i$ in (\ref{eq:formulaF}) are Grassmann-odd variables.

\section{Derivation of $\bm{G}$ (\ref{eq:G=vGv+vGv})}
\label{sec:G}

In this appendix, we summarize the derivation of the expression
(\ref{eq:G=vGv+vGv}) for $G$ from the original one (\ref{eq:Gagain}).
First, each term on the RHS of (\ref{eq:Gagain}) is calculated to
give the following form not containing $M_\pm$:
\begin{align}
&(\vvmz-\vvpz,\vvpz)(1-T\calM)^{-1} T
\Pmatrix{\vvpz-\vvmz \\ \vvmz}
\nn\\
&\quad
=-(\vvpz-\vvmz)D^{-1} T(\vvpz-\vvmz)
+\vvpz D^{-1}T(1-T)\vvmz ,
\label{eq:G2nd}
\\[2ex]
&2\,(\vvmz-\vvpz,\vvpz)
(1-T\calM)^{-1}\Pmatrix{0\\ \bm{t}}
\nn\\
&\quad
=2\,\vvpz (1-T) D^{-1}\bm{t}
=-2\,\vvpz D^{-2}(1-T)^3\left(\vvpz + \vvmz\right) ,
\label{eq:G3rd}
\\[2ex]
&(0,\bm{t})\calM (1-T\calM)^{-1}\Pmatrix{0\\ \bm{t}}
\nn\\
&\quad
=\bm{t} D^{-1}M_0(1-T)\bm{t}
=(\vvpz+\vvmz)D^{-3}M_0(1-T)^5 (\vvpz+\vvmz) .
\label{eq:Glast}
\end{align}
In every calculation we have used (\ref{eq:new(1-TM)^-1}) for
$(1-T\calM)^{-1}$ and defined
\begin{equation}
D=(1-M_0)(1+T) .
\label{eq:D}
\end{equation}
We also made repeated use of the identities (\ref{eq:M+M+M=1}),
(\ref{eq:MvI}) and (\ref{eq:MvII}).
In particular, at the first equality of (\ref{eq:G3rd}) we have used
\begin{equation}
\vvpz M_- + \vvmz M_+=0 ,
\label{eq:vMI}
\end{equation}
obtained by taking the transpose of (\ref{eq:MvI}) and using that
$M_\alpha$ is a symmetric matrix
\begin{equation}
M_\alpha^\T=M_\alpha , \qquad (\alpha=0,\pm).
\label{eq:M^T=M}
\end{equation}
In obtaining the last expressions of (\ref{eq:G3rd}) and
(\ref{eq:Glast}), we have used (\ref{eq:t}) for $\bm{t}$.

Then, summing (\ref{eq:G2nd})--(\ref{eq:Glast}) and the LHS of
(\ref{eq:vv=V00}), and using that $\vvmz\calO\vvpz=\vvpz\calO\vvmz$
and $\vvmz\calO\vvmz=\vvpz\calO\vvpz$ when $\calO$ consists solely of
$M_0$ and $T$ commutative with $C$, we get primitive
expressions of $G_{++}$ and $G_{+-}$. The final expressions
(\ref{eq:G++}) and (\ref{eq:G+-}) are obtained by substituting
\begin{equation}
D^{-1}=\left[(1-M_0)(1+3 M_0)\right]^{-1}
\left(1+2 M_0 - M_0 T\right) ,
\label{eq:D^-1}
\end{equation}
combining all terms over a common denominator,
and repeatedly using $M_0 T^2=(1+M_0)T-M_0$
(see (\ref{eq:M0T^2-(1+M0)T+M0=0})) to reduce the power of $T$
in the numerator.

\section{Derivation of $\bm{H}$ (\ref{eq:H=vHv+vHv})}
\label{sec:H}

The main task in obtaining the concise expression (\ref{eq:H=vHv+vHv})
for $H$ is the calculation of the first three terms in the exponent
in the last expression of (\ref{eq:KphitphitphitV}):
\begin{equation}
-\frac12\bm{V}(1-S\calV_3)^{-1}S\bm{V}
+\bm{V}(1-S\calV_3)^{-1}\bm{t} A_0
-\frac12 A_0\bm{t}\calV_3(1-S\calV_3)^{-1}\bm{t} A_0
-\frac12 V_{00}(A_0)^2
\label{eq:exponent}
\end{equation}
Let us first evaluate these three terms.

\noindent
\underline{
$-(1/2)\bm{V}(1-S\calV_3)^{-1}S\bm{V}$}

First, we have
\begin{align}
&(1-S\calV_3)^{-1}=(1-T\calM_3)^{-1}
\nn\\
&\quad
=D_3^{-1}
\Pmatrix{
(1-M_0)(1+T)&T(M_+ +M_-T)&T(M_- +M_+T)\\
T(M_- +M_+T)&(1-M_0)(1+T)&T(M_+ +M_-T)\\
T(M_+ +M_-T)&T(M_- +M_+T)&(1-M_0)(1+T)
} ,
\label{eq:(1-SV3)^-1}
\end{align}
where $D_3$ is the $3\times 3$ determinant of $1-S\calV_3$,
(\ref{eq:det3x3}):
\begin{equation}
D_3=1 - 3\,M_0 + 3\,T -T^3 .
\label{eq:D3}
\end{equation}
Using this and $\bm{V}$ in (\ref{eq:newquantities}), the first term of
(\ref{eq:exponent}) is calculated as
\begin{align}
&-\frac12\bm{V}(1-S\calV_3)^{-1}S\bm{V}
=-\frac12\bm{V}(1-S\calV_3)^{-1}T\,C\bm{V}
\nn\\
&
=-\frac12\,(\vvpz,\vvmz)
\Pmatrix{a^{(2)}_0&a^{(3)}_0&a^{(1)}_0\\a^{(3)}_0&a^{(1)}_0&a^{(2)}_0}
(1-S\calV_3)^{-1}\,T
\Pmatrix{a^{(3)}_0& a^{(2)}_0\\
         a^{(1)}_0& a^{(3)}_0\\
         a^{(2)}_0& a^{(1)}_0}\Pmatrix{\vvpz\\ \vvmz}
\nn\\
&=\frac14(A_0)^2\,(\vvpz,\vvmz) D_3^{-1} T
\nn\\
&\qquad
\times\Pmatrix{
(1-T)(1-T+ 3TM_+)&
(1+T)(-2+T)(1-M_0)\\
(1+T)(-2+T)(1-M_0)&
(1-T)(1-T+ 3TM_-)
}\!
\Pmatrix{\vvpz\\ \vvmz} .
\label{eq:pause}
\end{align}
In the last line we have replaced $M_\mp$ in the
upper-left/lower-right component of the $2\times 2$ matrix
by $1-M_0 -M_\pm$ via (\ref{eq:M+M+M=1}).
This is because we can remove the remaining $M_\pm$ in
the matrix by using (\ref{eq:MvII}), i.e.,
$M_\pm\bm{v}_{\pm 0}=\bm{v}_{\pm 0}-M_0\bm{v}_{\mp 0}$.
After carrying out this procedure, we finally obtain
\begin{align}
-\frac12\bm{V}(1-S\calV_3)^{-1}S\bm{V}
=&\frac14(A_0)^2\,(\vvpz,\vvmz) D_3^{-1}
T(1-T)\Pmatrix{
1+2T&-2 -T\\-2 -T&1+2T}\!
\Pmatrix{\vvpz\\ \vvmz} .
\label{eq:-1/2V(1-SV3)^-1SV}
\end{align}

\noindent\underline{
$\bm{V}(1-S\calV_3)^{-1}\bm{t} A_0$}

Eq.\ (\ref{eq:(1-SV3)^-1}) together with (\ref{eq:simpleeqT}),
(\ref{eq:CsimpleeqT}) and (\ref{eq:t}), i.e.,
\begin{equation}
(1-M_0)(1+T)\bm{t}=
-(1-T)^2\left(\vvpz+\vvmz\right) .
\label{eq:tIagain}
\end{equation}
gives
\begin{align}
&(1-S\calV_3)^{-1} \bm{t}A_0
=-D_3^{-1}
(1-T)
\nn\\
&\quad\times
\Pmatrix{
(1-T)(\vvpz + \vvmz)&T(\vvpz-T\vvmz)
&T(\vvmz-T\vvpz)\\
T(\vvmz-T\vvpz)&(1-T)(\vvpz +\vvmz)
&T(\vvpz-T\vvmz)\\
T(\vvpz-T\vvmz)&T(\vvmz-T\vvpz)
&(1-T)(\vvpz +\vvmz)
}A_0
\nn\\
&
=-D_3^{-1}
(1-T)
\nn\\
&\quad\times
\Pmatrix{
(1-T)a^{(1)}_0 +T a^{(2)}_0 -T^2 a^{(3)}_0&
(1-T)a^{(1)}_0 +T a^{(3)}_0 -T^2 a^{(2)}_0\\
(1-T)a^{(2)}_0 +T a^{(3)}_0 -T^2 a^{(1)}_0&
(1-T)a^{(2)}_0 +T a^{(1)}_0 -T^2 a^{(3)}_0\\
(1-T)a^{(3)}_0 +T a^{(1)}_0 -T^2 a^{(2)}_0&
(1-T)a^{(3)}_0 +T a^{(2)}_0 -T^2 a^{(1)}_0
}\Pmatrix{\vvpz\\ \vvmz} .
\label{eq:(1-SV3)^-1tA0}
\end{align}
The second term of (\ref{eq:exponent}) is easily calculated by
using (\ref{eq:(1-SV3)^-1tA0}):
\begin{align}
\bm{V}(1-S\calV_3)^{-1}\bm{t} A_0
&=\frac12(A_0)^2\, (\vvpz,\vvmz)D_3^{-1} (1-T)
\nn\\
&\qquad\times
\Pmatrix{
1-3T-T^2 & 1 + 2T^2 \\
1 + 2T^2     & 1-3T-T^2}
\Pmatrix{\vvpz\\ \vvmz} .
\label{eq:V(1-SV3)^-1tA0}
\end{align}

\noindent\underline{
$-(1/2)A_0\bm{t}\calV_3(1-S\calV_3)^{-1}\bm{t} A_0$}

Let us calculate
\begin{align}
-\frac12\,A_0\bm{t}\calV_3(1-S\calV_3)^{-1}\bm{t} A_0
=-\frac12\,A_0\bm{t}\calM_3(1-S\calV_3)^{-1}\bm{t} A_0 .
\label{eq:-1/2A0tV3(1-SV3)^-1tA0pre1}
\end{align}
First, using (\ref{eq:(1-SV3)^-1tA0}) we have
\begin{align}
&-A_0\calM_3(1-S\calV_3)^{-1}\bm{t} A_0
\nn\\
&=\frac12\,(A_0)^2 D_3^{-1} (1-T)\sum_{\pm}
\left((2-3T+T^2)M_0-(1+2T^2)M_\pm
+(-1+3T+T^2)M_\mp\right)\bm{v}_{\pm 0}
\nn\\
&
= -\frac12\,(A_0)^2 D_3^{-1}(1-2T)(1-T)^2\,(\vvpz + \vvmz) .
\label{eq:-1/2A0tV3(1-SV3)^-1tA0pre2}
\end{align}
In obtaining the last line we have used the same technique as used
in passing from (\ref{eq:pause}) to (\ref{eq:-1/2V(1-SV3)^-1SV}).
As $\bm{t}$ to be multiplied to
(\ref{eq:-1/2A0tV3(1-SV3)^-1tA0pre2}) to get
(\ref{eq:-1/2A0tV3(1-SV3)^-1tA0pre1}), we use the following
expression:
\begin{align}
\bm{t}&= -D^{-1}(1-T)^2(\vvpz + \vvmz)
\nn\\
&=-(1+3M_0)^{-1}(1+T)(\vvpz + \vvmz) ,
\label{eq:tanother}
\end{align}
where we have used (\ref{eq:D^-1}) and the identity
\begin{equation}
\left(1+ 2M_0 -M_0T\right)(1-T)^2
=(1-M_0)(1+T) ,
\end{equation}
which is obtained by repeated use of (\ref{eq:M0T^2-(1+M0)T+M0=0}).
Then, we have
\begin{align}
-\frac12\,A_0\bm{t}\calV_3(1-S\calV_3)^{-1}\bm{t} A_0
=&\frac14 (A_0)^2\,(\vvpz+\vvmz) (1+3M_0)^{-1} D_3^{-1}
\nn\\
&\quad\times
(1-2T)(1-T)^2 (1+T)(\vvpz+\vvmz) .
\label{eq:-1/2A0tV3(1-SV3)^-1tA0}
\end{align}

\noindent\underline{Total of $H$}

Now we have obtained the first three terms in (\ref{eq:exponent}),
and the total of $H$ is given by
\begin{equation}
H=\left(-\frac12 (A_0)^2\right)^{-1}
\Bigl[\mbox{(\ref{eq:-1/2V(1-SV3)^-1SV})}
+\mbox{(\ref{eq:V(1-SV3)^-1tA0})}
+\mbox{(\ref{eq:-1/2A0tV3(1-SV3)^-1tA0})}
\Bigr]+ V_{00} + \bm{t}(1+T)^{-1}\bm{t} .
\label{eq:totalH}
\end{equation}
We use (\ref{eq:vv=V00}) for $V_{00}$ and
\begin{equation}
\bm{t}(1+T)^{-1}\bm{t}
=(\vvpz+\vvmz)(1+3 M_0)^{-2}(1+T)(\vvpz+\vvmz) ,
\label{eq:t(1+s)^-1t}
\end{equation}
obtained from (\ref{eq:tanother}).
Then, using
\begin{equation}
D_3^{-1}=\left[(1-M_0)^2(1+ 3M_0)\right]^{-1}
\left(1+M_0-M_0^2 -M_0 T\right) ,
\end{equation}
in (\ref{eq:-1/2V(1-SV3)^-1SV}), (\ref{eq:V(1-SV3)^-1tA0}) and
(\ref{eq:-1/2A0tV3(1-SV3)^-1tA0}), and putting all terms in
(\ref{eq:totalH}) over a common denominator, and reducing the power of
$T$ by repeated use of (\ref{eq:M0T^2-(1+M0)T+M0=0}) as we did for
$G$, we arrive at (\ref{eq:H=vHv+vHv}).

\end{document}